\documentclass[preprint,11pt]{JHEP3}

\JHEPspecialurl{http://jhep.sissa.it/JOURNAL/JHEP3.tar.gz}

\usepackage{epsfig,multicol,amsmath}

\usepackage{epstopdf}
\usepackage{bm}  
\usepackage{amsmath}     
\usepackage{epsfig,epsf}
\usepackage{graphicx}
\usepackage{rotating}
\usepackage{cite}

\def\beq{\begin{equation}}
\def\eeq{\end{equation}}
\def\bea{\begin{eqnarray}}
\def\eea{\end{eqnarray}}

\def\eq#1{{Eq.~(\ref{#1})}}
\def\fig#1{{Fig.~\ref{#1}}}
\newcommand{\bas}{\bar{\alpha}_S}
\newcommand{\as}{\alpha_S}

\newcommand{\Lb}{\left(}
\newcommand{\Rb}{\right)}

\parskip=\itemsep               

\newcommand{\nn}{\nonumber}

\title{Violation of the geometric scaling behaviour of the amplitude for running QCD coupling in the saturation region.}
\author{\Large 
Bastian Diaz Saez${}^a$ \thanks{Email: $\mbox{bastilo}_{\_}1$@hotmail.com}\, and\,Eugene\, Levin${}^{a, b}$ \thanks{Email: leving@post.tau.ac.il, eugeny.levin@usm.cl}
\\
${}^a$\, Departamento de F\'\i sica, 
Centro de Estudios Subat$\acute{o}$micos,
Universidad T$\acute{e}$cnica Federico Santa Mar\'\i a,\\ and
Centro Cient´ıfico-Tecnol$\acute{o}$gico de Valpara\'\i so,
Casilla 110-V,  Valparaiso, Chile\\
${}^b$ \, Department of Particle Physics, School of Physics and Astronomy,
Tel Aviv University, Tel Aviv, 69978, Israel\\
}

\abstract
{In this paper we show that the intuitive guess that the geometric scaling behaviour should be violated in the case of the running QCD coupling, turns out to be correct. The scattering amplitude of the dipole with the size $r$  depends on new dimensional scale: $\Lambda_{QCD}$,  even at large values $Y = \ln(1/x)$ and $l\,=\,
\ln\Lb \as\Lb r^2\Rb/\as\Lb 1/Q^2_s\Rb\Rb$. However, in this region we found a new scaling behaviour: the amplitude is a function of $\zeta = Y\,l$.
We state  that only  in the vicinity of the saturation scale $Q_s$ ($\as(Q^2_s)\,\ln \Lb r^2 Q^2_s\Rb\,\leq \, 1$),  the amplitude shows the geometric scaling behaviour.
Based on these finding the geometric scaling behavior  that has been seen experimentally, stems from either we have not probed the proton at HERA and the LHC  deeply inside the saturation region or that there exists the mechanism of  freezing of the QCD  coupling constant at $r^2 \approx 1/Q^2_s$.

}
\keywords{Colour Glass Condensate, gluon saturation,  non-linear evolution, geometric scaling behaviour
}
\dedicated{PACS: 12.38-t, 12.38.Cy,1 2.38.Lg, 13.60.Hd, 24.85.+p, 25.30.Hm}
\preprint{TAUP 2920/11 \\
{\tt }\\
\today}

\begin{document}

\section{Introduction.}
The geometric scaling behaviour of the scattering amplitude gives an example of the prediction that based on the most fundamental
features of the high parton density QCD. It means that the scattering amplitude of the dipole in the saturation region is a function of one dimensionless variable $\tau\,\,=\,\,r^2\,Q_s^2\Lb Y; b\Rb$ instead of being a function of dipole size ($r$), energy ($ Y = \ln(1/x)$) and impact parameter ($b$).  $Q_s(x,b)$ is the new  dimensional scale ( saturation momentum) which absorbed entire dependence on energy and impact parameter of the amplitude. The existence of this scale and its appearance  from the non-linear evolution at low $x$ is the most fundamental theoretical result of both the BFKL Pomeron calculus\cite{BFKL,GLR,MUQI,BRN,BART} and the Colour Glass Condensate (CGC) 
approach \cite{MV,JIMWLK,B,K}.   The idea of the geometric scaling behaviour  is very simple: $\tau$ is the only dimensionless variable in the dense system of partons. At the moment we have a general proof of the geometric scaling behaviour of the amplitude in the saturation region\cite{BALE}, two examples of the analytically solved non-linear Balitsky-Kovchegov equation with simplified kernels\cite{LETU,MUPE}
which show the geometric scaling behaviour and the proof that this behaviour is a general property of the linear dynamics in the vicinity of the saturation scale\cite{IIM}.

However, analyzing these arguments and proofs one can see that they are related to the case of fixed QCD coupling constant
as far as the behaviour of the amplitude in the saturation domain is concerned.
 Indeed, for frozen QCD coupling we do not have other dimensional parameters  but $Q_s$. For running QCD coupling the situation is not so clear since 
it brings the second domensional scale:    $ \Lambda_{QCD}$, and the role of this scale has to be studied in the  saturation region.

The paper presents such study in the case of the simplified BFKL kernel (see Ref.\cite{LETU}).  We show that the geometic scaling behaviour is violated for the running QCD coupling  and the amplitude does not depend on the one variable $\tau$. It turns out that the amplitude depends on a different variable
\beq \label{ZETA}
\zeta\,=\,\frac{4 N_c}{b}\, Y\ln\Lb\frac{ \bas\Lb r^2 \Lambda^2_{QCD}\Rb}{\bas\Lb \Lambda^2_{QCD}/Q^2_s\Rb}\Rb\,\,\,\,\,\,\,
\mbox{where} \,\,\,\,\,\,\,\bas (r^2)\,= \,\frac{4 N_c}{b \,\ln\Lb1/\Lb r^2 \Lambda_{QCD}^2\Rb\Rb}
\eeq
 with  $b = 11 N_c /3 - 2 N_f/3$ for number of colours $N_c$ and the number of flavours $N_f$. One can see that $\zeta$ depends on both dimemsional scales: $Q_s$ and $\Lambda_{QCD}$.

\section{General approach: behaviour of the scattering amplitude in the vicinity of the saturation scale for running QCD coupling}

The nonlinear  Balitsky-Kovchegov equation for the scattering amplitude of the dipole with size $r$ has the following form\cite{B,K}: 
\bea 
\frac{\partial N\Lb r,Y;b \Rb}{\partial\,Y}\,
\,\,&
=&\,\,\int\,\frac{d^2 r_1}{2 \pi}\,K\Lb r; r_1,r_2\Rb \times 
\Big\{N\Lb r_1,Y;\vec{b} \,-\, 
\frac{1}{2}\,\vec{r}_2\Rb \,+\, N\Lb r_2,Y;\vec{b} \,-\, 
\frac{1}{2}\,\vec{r}_1\Rb\,-\, \,\, N\Lb r,Y;\vec{b}\Rb\,\,\nn\\
 &-& \,\,
 N\Lb r_1,Y;\vec{b} - \frac{1}{2}\,\vec{r}_2 \Rb\, N\Lb r_2 ,Y;\vec{b}- \frac{1}{2} \vec{r}_1\Rb \Big\} \label{GA1}
\eea
where $Y = \ln(1/x)$ is the rapidity of the incoming dipole; $N$ is the imaginary part of the scattering amplitude and $b$ is the impact parameter of this scattering process and $\vec{r}_2 = \vec{r} - \vec{r}_1$. The BFKL kernel $K\Lb r_1,r_2\Rb$ has the following form
\beq \label{K}
 K\Lb r; r_1,r_2\Rb\,\,=\,\,\bas\Lb r^2\Rb \left\{ \frac{r^2}{r^2_1\,r^2_2}\,\,+\,\,\frac{1}{r_1^2}\Lb\frac{\bas\Lb r^2_1\Rb}{\bas\Lb r^2_2\Rb}\,-\,1\Rb
\,+\,\frac{1}{r_2^2}\Lb \frac{\bas\Lb r^2_2\Rb}{\bas\Lb r^2_1\Rb}\,-\,1\Rb\right\}
\eeq
This kernel takes into account the running QCD coupling and was derived in Re.\cite{RAL}.
In \eq{GA1}
$\as $ is the QCD coupling 
\beq \label{AL}
\as\Lb r^2\Rb\,\,=\,\,\frac{\as\Lb R^2\Rb}{1\,\,+\,\,\frac{\as\Lb R^2\Rb}{4 \pi b}\ln\Lb R^2/r^2\Rb}\,\,=\,\,\frac{4 \pi}{b \,\ln\Lb 1/\Lb r^2\,\Lambda_{QCD}\Rb\Rb}
\eeq
 and  $\bas = N_c \as/\pi$. $R$ is the arbitrary size (so called the renormalization point)  which the physical observables do not depend on.

In the vicinity of the saturation scale where $r^2 \, \approx r^2_1\, \approx r^2_2\, \approx 1/Q^2_s$  and we can consider  that $\bas\Lb r^2\Rb \,=\,\bas\Lb r^2_1\Rb = \bas \Lb r^2_2\Rb$. Indeed, choosing $R = r$ we can see that 
\beq \label{GA2}
\as\Lb r^2_i \Rb\,\,=\,\,\frac{\as\Lb r^2\Rb}{1\,\,+\,\,\frac{\as\Lb r^2\Rb}{4 \pi b}\ln\Lb r^2/r^2_i\Rb}\,\,\,\,\xrightarrow{ \ln\Lb r^2/r^2_i\Rb\,\ll\,ln\Lb r^2 \,\Lambda_{QCD}\Rb }\,\,\,\,\as\Lb r^2 \Rb
\eeq
In the vicinity of the saturation scale $r^2 \,\varpropto \,1/Q^2_s$ and condition  $|\ln\Lb r^2_i\,Q^2_s\Rb|\,\ll\,\ln \Lb Q^2_s/\Lambda_{QCD}\Rb$ determines the  kinematic region which we call vicinity of the saturation scale. Using this simplification the kernel of \eq{GA1} looks as follows:
\beq \label{KK}
K\Lb r; r_1,r_2\Rb\,\,\,=\,\,\bas\Lb r^2\Rb \,\frac{r^2}{r^2_1\,r^2_2}
\eeq

The second simplification stems from the observation that for the equation for the saturation scale we do not need to know the precise  form of non-linear term \cite{GLR,MUTR,MUPE}. Therefore, to find this equation as well as behaviour of the amplitude in the vicinity of the saturation scale we need to solve the linear BFKL equation, but in the way which will be suitable for the solution of the non-linear equation with a general non-linear term. It is enough to use the semiclassical approximation for the amplitude $N\Lb r,Y;b \Rb$, which has the form
\beq \label{SCSOL}
N_A\Lb Y, \xi\Rb\,\,= \,\,e^{S\Lb Y, \xi\Rb}\,\,=\,\,e^{\omega\Lb Y,\xi\Rb\,Y\,+\,\Lb 1 - \gamma\Lb Y; \xi\Rb\Rb\,\xi\,+\,S_0} 
\eeq
where $\xi \,=\,\ln\Lb r^2 Q^2_s\Lb Y = Y_0; b \Rb\Rb$. In \eq{SCSOL} we are searching  for functions  $\omega\Lb Y,\xi\Rb$ and $\gamma\Lb Y,\xi\Rb$
which are smooth functions of  both arguments in the following sense
\beq \label{GA3}
\omega'_{Y}\Lb Y,\xi\Rb\,\ll\,\omega\Lb Y,\xi\Rb;\,\,\,\,\omega'_{\xi}\Lb Y,\xi\Rb\,\ll\,\omega\Lb Y,\xi\Rb;\,\,\,\,\gamma'_{Y}\Lb Y,\xi\Rb\,\ll\,\gamma\Lb Y,\xi\Rb;\,\,\,\,\gamma'_{\xi}\Lb Y,\xi\Rb\,\ll\,\gamma\Lb Y,\xi\Rb;
\eeq

The BFKL equation near to the saturation scale looks as follows
\beq \label{GA4}
\frac{\partial N\Lb r,Y;b \Rb}{\partial\,Y}\,
\,\,
=\,\,\bas\Lb r^2\Rb \int\,\frac{d^2 r_1}{2 \pi}\,K\Lb r; r_1,r_2\Rb \times 
\left\{N\Lb r_1,Y;\vec{b} \Rb \,+\, N\Lb r_2,Y;\vec{b}  \Rb\,-\, \,\, N\Lb r,Y;\vec{b}\Rb\,\right\}
\eeq
In \eq{GA4} we assume that we are looking for the solution at $b \,\gg\,r_1$ or/and $r_2$. Substituting \eq{SCSOL} into \eq{GA4} and taking into account that function $(r^2)^f \equiv \exp\Lb f \,\xi\Rb$ is the eigenfunction of the BFKL equation,  namely,
\bea \label{GA5}
&&\bas\Lb r^2\Rb \int\,\frac{d^2 r_1}{2 \pi}\,K\Lb r; r_1,r_2\Rb\,(r^2_1)^f \,\,=\,\, \bas\Lb r^2\Rb\chi\Lb f \Rb\,(r^2)^f\,\,
\mbox{with}\,\,\,
\chi\Lb f \Rb \,\,=\,\,2 \,\psi(1) - \psi(f) - \psi(1 - f) \nn\\
&&\,\,\mbox{where} \,\,\, \psi(z) = d \ln \Gamma(z)/d z\,\,\,\,\,  \mbox{ and } \,\,\,\,\,\,\Gamma(z) \,\,\,\, \mbox{ is Euler gamma function}
\eea
we obtain that
\beq \label{GA6}
\omega\Lb Y,\xi\Rb\,\,\,=\,\,\bas\Lb \xi\Rb\,\chi\Lb \gamma\Lb Y, \xi\Rb\Rb
\eeq

This solution has a form of wave-package and the critical line is the specific trajectory for this wave-package which coincides with the its front line. In other words, it is the  trajectory on which the phase velocity ($v_{ph}$) for the wave-package is the same as
the group velocity ( $v_{gr}$). The equation $v_{gr} \,\,=\, v_{ph}$ has the folowing form for \eq{GA6}
\beq \label{CRL}
v_{ph}\,\,=\,\,\bas
\Lb r^2\Rb \frac{\chi\Lb \gamma_{cr} \Rb}{ 1 \,-\,\gamma_{cr}}\,\,\,=\,\,\,- \bas\Lb r^2 \Rb \chi'\Lb \gamma_{cr}\Rb\,\,=\,\,v_{gr}
\eeq
with the solution $\gamma_{cr} \,=\,0.37$.

\eq{CRL} can be translated into the following equation for the critical trajectory
\beq \label{GA7}
\frac{d \xi\Lb Y\Rb}{ d Y}\,\,=\,\,v_{ph}\,\,=\,\,\bas\Lb \xi\Rb\,\frac{\chi\Lb \gamma_{cr} \Rb}{ 1 \,-\,\gamma_{cr}}
\eeq
with the solution
\beq \label{GA8}
\frac{8 N_c}{b}\,\,\frac{\chi\Lb \gamma_{cr} \Rb}{ 1 \,-\,\gamma_{cr}}\,Y\,\equiv\,\,\xi^2_s\,\,=\,\,\xi^2\,\,-\,\,\xi^2_0\,\,\
\eeq
where $\,\,\,\xi_0\,=\,\ln\Lb  Q^2_s\Lb Y = Y_0; b \Rb/\Lambda^2_{QCD}\Rb$   and   $\xi_s \,=\,\ln\Lb Q^2_s(Y,b)/Q_s^2(Y=Y_0,b)\Rb$

For finding the behaviour of the amplitude in the vicinity of  the line given by \eq{GA8} one should expand function $\omega\Lb Y,\xi\Rb$ and $\gamma\Lb Y; \xi\Rb$ and find a deviation from the critical line of \eq{GA8}. Replacing $\xi = \xi_s \,+\,\Delta \xi$ where $\Delta \xi \,=\,\ln\Lb 
r^2\,Q^2_s\Lb  Y,b\Rb\Rb$ and considering $\Delta \xi \,\ll\,\xi_s$ one obtain
\bea \label{GA9}
N\Lb Y,\xi\Rb \,\,\,&\varpropto &\,\,\,\exp\left\{ \Lb \frac{\partial \omega\Lb Y, \xi = \xi_s\Rb}{\partial \xi}\,Y\,+\,1 - \gamma\Rb\,\Delta \xi\right\}\nn\\
                            & = &\,\, \exp\left\{ \Lb \frac{\chi\Lb \gamma_{cr}\Rb}{\xi^2_s}\,Y + 1 - \gamma_{cr}\Rb \Delta \xi\right\}\nn\\
      &=&\,\,\, \Big( r^2 Q^2_s\Lb Y, b\Rb\Big)^{  \frac{3}{2}\Lb 1 - \gamma_{cr}\Rb}
\eea

It should be mentioned that everything, except \eq{GA9},  are  not new and have been studied in details before (see for example Refs.\cite{GLR,MUTR,MUPE}). We discuss them here for the complitness of presentation. \eq{GA9} shows that the running QCD coupling leads to 
a different behaviour of the scattering amplitude in the vicinity of the critical trajectory. Recall that for frozen $\bas$ the amplitude  $N \,\varpropto \Big( r^2 \,Q^2_s\Big)^{-(1 - \gamma_{cr})}$. Concluding this section we would like to stress that we obtain the geometric scaling behaviour of the scattering amplitude to the right of the critical curve ($ \tau \,>\,1$) in the case of the running $\as$. This result gives us a hope that inside the saturation region we can observe the geometric scaling behaviour as well.

\begin{boldmath}
\section{The non-linear equation with the simplified  BFKL kernel for running $\as$.}
\end{boldmath}
The solution to the Balitsky-Kovchegov equation with the kernel of \eq{K} has not been found.  Following Ref. \cite{LETU} we simplify the kernel by taking into account only log contributions. In other words, we would like to consider only leading twist contribution   to the BFKL kernel, which contains all twists.
Actually we have two types of the logarithmic contributions:  $\ln\Lb r^2 \Lambda^2_{QCD}\Rb$ for $r^2 \,\ll\, 1/Q^2_s$ and $  \ln\Lb r^2\,Q^2_s\Rb$ for   $r^2 \,>\,1/Q^2_s$.

\begin{boldmath}
\subsection{$r^2 \,\ll\, 1/Q^2_s$}
\end{boldmath}

 In this kinematic region we can simplify  $K\Lb r;r_1,r_2\Rb$ in \eq{K}  in the following way\cite{LETU}, since $r_1
 \gg r$ and $r_2=|\vec{r} - \vec{r}'| > r$
 \beq \label{K1}
 \int d^2 r' \,K\Lb r, r_1.r_2\Rb\,\,\rightarrow\,\pi\,\bas\Lb r^2 \Rb  r^2\,\int^{\frac{1}{\Lambda^2_{QCD}}}_{r^2} \frac{ d r'^2}{r'^4}
\eeq
Introducing $\tilde{N}\Lb r,Y;b \Rb= N\Lb r,Y;b\Rb/\Lb \bas\Lb r^2\Rb\,r^2\Rb$ we obtain
 \beq \label{SK1}
\frac{\partial \tilde{N}\Lb r, Y;b\Rb}{\partial Y}\,\,=\,\,\int^{1/\Lambda_{QCD}^2}_{r^2}\,d r'^2\,\left\{\frac{\bas\Lb r'^2\Rb}{r'^2}\,\tilde{N}\Lb r', Y;b\Rb\,-\,\frac{\bas^2\Lb r'^2\Rb}{2} \tilde{N}^2\Lb r',Y;b\Rb\right\}
\eeq

One can see that  the simplified kernel of \eq{K1} sums $\Lb\int^{1/\Lambda_{QCD}^2}_{r^2}\,d r'^2\,\frac{\bas\Lb r'^2\Rb}{r'^2}\Rb^n$. As we have discussed in the previous section the form of non-linear corrections is not important here. One can see that the linear part of \eq{SK1} 
gives the familiar GLAP equation in the double log approximation\cite{DGLAP}. 

\begin{boldmath}
\subsection{$r^2 \,\gg\, 1/Q^2_s$}
\end{boldmath}

The main contribution in this kinematic region originates from the decay of the large size dipole into one small size dipole  and one large size dipole.  However, the size of the small dipole is still larger than $1/Q_s$.  It turns out that
$\bas$
 depends on the size of produced dipole if this size is the smallest one. It follows directly from \eq{K} in the kinematic regions: 
$ r \approx r_2 \,\gg\,r_1\,\,\gg\,1/Q^2_s$  and $ r \approx r_1 \,\gg\,r_2\,\,\gg\,1/Q^2_s$
 (see
Ref. \cite{MURC} for additional arguments). This observation can be translated in the following form of the kernel
\beq \label{K2}
 \int d^2 r' \,K\Lb r, r'\Rb\,\,\rightarrow\,\pi\, \int^{r^2}_{1/Q^2_s(Y,b)} \frac{\bas\Lb 
r^2_1\Rb d r^2_1}{r^2_1}\,\,+\,\,
\pi\, \int^{r^2}_{1/Q^2_s(Y,b)} \frac{ \bas\Lb r^2_2\Rb d r^2_2}{r^2_2}
\eeq

One can see that this kernel leads to the $\Lb  \int^{r^2}_{1/Q^2_s(Y,b)} \frac{\bas\Lb r^2_1\Rb d r^2_1}{r^2_1}\Rb^n$-contributions. Introducing a new function
\beq \label{TN}
 \tilde{N} \Lb r,Y;b\Rb\,\,=\,\,\int^{r^2}_{1/Q^2_s} d r'^2\,\frac{\bas\Lb r'^2\Rb}{r'^2} \,N\Lb r',Y;b\Rb
\eeq
one obtain the following equation
\beq \label{SK2}
\frac{\partial N\Lb r, Y; b\Rb}{\partial Y}\,\,=\,\, \tilde{N} \Lb r,Y;b\Rb\,\Big( 1\,\,-\,\,N\Lb r, Y; b\Rb \Big)
\eeq
Introducing a new variable
\beq \label{L}
l\,\,=\,\,\int^{r^2}\,d r'^2\frac{\bas\Lb r'^2\Rb}{r'^2}\,\,=\,\,\frac{4 N_c}{b}\,\ln\Big(1/  \bas\Lb r^2 \Rb\Big)\,\,=\,\,\frac{4 N_c}{b}\,\ln \Lb \bar{\xi}\Rb
\eeq
with $\bar{\xi}\,=\,-\ln\Lb r^2\,\Lambda^2_{QCD}\Rb\,\equiv\,- \xi$
and new function $\phi\Lb r, Y; b \Rb$
\beq \label{PHI}
N\Lb r, Y; b\Rb\,\,=\,\,1\,\,\,-\,\,e^{- \phi\Lb r,Y;b\Rb}
\eeq
we obtain the following equation
\beq \label{SK3}
\frac{ \partial^2 \phi\Lb r, Y; b \Rb}{\partial Y \partial l}\,\,=\,\,1\,\,\,-\,\,\,e^{ - \phi\Lb r,Y;b\Rb}
\eeq

\section{Solutions to the simplified equation}
\subsection{Initial and boundary conditions. }
The simplified BFKL kernel looks  as follows \cite{LETU} in $\omega$ and $\gamma$ representation (in double Mellin transform with respect to $Y$ and $\xi$)
\bea \label{KSM}
\chi\Lb \gamma\Rb\,\,=\,\,\left\{\begin{array}{l}\,\,\,\frac{1}{\gamma}\,\,\,\,\,\mbox{for}\,\,\,r^2\,\leq \,1/Q^2_s;\\ \\
\,\,\,\frac{1}{1 \,-\,\gamma}\,\,\,\,\,\mbox{for}\,\,\,r^2\,>\,1/Q^2_s\,; \end{array}
\right.
\eea

Using this kernel for small values of $r^2$ ( $ r^2 < 1/Q^2_s$)   and the general formulae of  \eq{GA8} and \eq{GA9} one can write the initial conditions at $ \tau = r^2\,Q^2_s\,=\,1$. They are
\beq 
\label{IBC}
\phi\Lb Y,\bar{\xi} = \xi_s; b\Rb\,=\,\phi_0;\,\,\,\,\,\,\,\,\,\,\,\frac{ \partial \phi\Lb Y, \bar{\xi} = \xi_s; b\Rb}{\partial \bar{\xi}}\,\,=\,\,-\,\frac{3}{4}\phi_0
\eeq

\begin{figure}
\begin{minipage}{100mm}{
\centerline{\epsfig{file=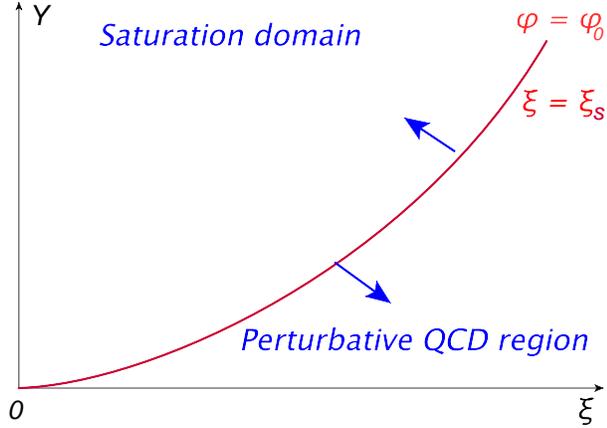,width=80mm}}
}
\end{minipage}
\begin{minipage}{70mm}{
\caption{ Kinematic regions for the  dipole scattering: $\xi =\,-\, \ln\Lb r^2\,\Lambda^2_{QCD}\Rb$ and $\xi_s \,=\,\ln\Lb Q^2_s/\Lambda^2_{QCD}\Rb$.  The  red line is the critical trajectory with the equation $ \frac{32 N_c}{b}\,Y\,\,=\,\,\xi^2$ (see \protect\eq{CR1})  on which $\phi\Lb Y, \xi\,=\,\xi_s\Rb = \phi_0$.}
\label{spic}
}
\end{minipage}
\end{figure}

The critical line that gives us the energy dependence of the saturation scale has the form (see \eq{GA8}
\beq \label{CR1}
\frac{32 N_c}{b}\,Y\,\,=\,\,\bar{\xi}^2
\eeq
In \eq{CR1} we assume that $\xi_0 = 0$. It means that we consider the scattering amplitude for the dipoles of all sizes smaller that $r_0 = 1/\Lambda_{QCD}$ and the entire kinematic region can be divided in two parts: the region of perturbative QCD and the saturation domain (see \fig{spic}).

\begin{boldmath}
\subsection{ Solution for $\phi\, \gg\, 1$ }
\end{boldmath}

Searching for the solution to \eq{SK3} we start  with  finding the asymptotic berhaviour of $\phi$ at large values of $Y$ and $l$. We expect that $\phi$ will be large in this region since that dipole amplitude tends to be close to unity due to unitarity constraints.  Therefore, in this region \eq{SK3}  degenerates to a very simple equation
\beq \label{PL1}
\frac{\partial^2 \phi\Lb Y,l;b\Rb}{\partial Y \partial l}\,\,=\,\,1
\eeq
with obvious solution:
\beq \label{PL2}
 \phi_{\infty}\Lb Y,l;b\Rb\,\,=\,\,Y\,l \,+\,F\Lb Y\Rb\,+\,G\Lb l \Rb
\eeq
where functions $F$ and $G$ should be found from the initial conditions of \eq{IBC}.  

The final solution has the form

\beq \label{PL5}
\tilde{\phi}_{\infty}\Lb Y, l;b\Rb\,\,=\,\,Y\,\Lb l \,-\,l_s\Rb \,-\,\frac{3}{4}\phi_0 \Lb e^{l} \,-\,e^{l_s}\Rb\,-\,\frac{1}{2} \Lb e^{2 l} \,-\,e^{2 l_s}\Rb\,\,+\,\,\phi_0
\eeq
where $l_s\,\,=\,\,\frac{ N_c}{b} \ln  \xi_s$.

Therefore, we learned two lessons in this subsection: (1)  the main problem with \eq{SK3} is to satisfy the initial and boundary conditions; and (2)
the asymptotic solution does not show a geometric scaling behaviour since even the simplest solution of \eq{PL5} does not depend on the variable $z = \xi_s - \bar{\xi}$. However, the solution of \eq{PL5} at $\bar{\xi }\,\to\, \xi_s$ has the following form
\beq \label{PL7}
\tilde{\phi}_{\infty}\Lb Y, l;b\Rb\,\,\xrightarrow{ \bar{\xi} - \xi_s \,\ll\,\xi_s}\,\,\phi_0 \,-\,\frac{3}{4} \phi_0 \Lb \bar{\xi} - \xi_s\Rb\,\,=\,\,\phi\,\,+\,\,\frac{3}{4}\,\phi_0 \,z
\eeq
showing the geometric scaling behaviour.  Hence, we can hope that the solution will show the geometric scaling behaviour in the vicinity of the saturation scale.

On the other hand, the solution  given by  \eq{PL5},        leads to $\phi_{\infty} \,<\,0$  and, therefore, contradicts the unitarity constraints, leading to the 
negative imaginary part of the amplitude.  Such behaviour stems from that terms in \eq{PL5}   which are responsible for the    matching of the $\partial \phi/\partial l$  at $l \to 0$.  Hence we have to find a diffrent solution which has the same behaviour  $ \phi_{\infty}  = Y \,l$  for $\phi \,\gg\,1$.

\subsection{Traveling wave solution }
\eq{SK3} has general traveling wave solution  (see Ref.\cite{MATH} formula {\bf 3.5.3})
which can be found noticing that  $\phi\Lb Y, l; b\Rb \,\equiv\,\phi\Lb \eta \equiv  a \,Y + b \,l;b\Rb$ reduced the equation to 
\beq \label{TW1}
a\,b \frac{d^2 \phi\Lb \eta; b \Rb}{d \eta^2}\,\,=\,\,1 - e^{- \phi\Lb \eta; b \Rb}
\eeq
The general solution of \eq{TW1} has the form
\beq \label{TW2}
\int^\phi_{\phi_0}\frac{d \phi'}{\sqrt{c \,+\,\frac{1}{2 \,a\,b}\Big( \phi'  - 1 + e^{-\phi'}\Big)}}\,\,=\,\, \eta\,\,=\,\, a\,Y \,+\,b \,l
\eeq
where $c, \phi_0, a$ and $b$ are arbitrary constants that  should be found from the initial and boundary conditions.

The initial conditions of \eq{IBC} can be written in terms of $Y$ and $l$ variables as
\beq \label{TWIC}
\phi\Lb \eta 
\,=\,a Y + b l_s; b \Rb \,\,=\,\,\,\phi_0\,;\,\,\,\,\,\,\,\,\,\,\,\,\phi'_\eta\Lb  \eta \,=\,a Y + bl_s; b \Rb\,\,=\,\,-\frac{3}{4}\,\phi_0 \,\xi_s
\eeq
It should be mentioned that the variable $\eta$ is not the scaling variable $z \,=\,\ln\Lb \tau\Rb \,=\,\xi_s - \bar{\xi}$ with $\xi_s = \sqrt{\frac{32 N_c}{b} \,Y}$.  One can see that we cannot satisfy the initial conditions  of \eq{TWIC}. 
Indeed, even to satisfy the first of \eq{TWIC} we need to choose $\eta = 0$ on the critical line. As you see we cannot do this with $a$ and $b$ being constants. The second equation depends on $Y$, but not on $\eta$, making impossible to satisfy this condition in the framework of traveling wave solution.

If we try to find a solution  which depends on $z$ ($\phi\Lb Y; r^2;b\Rb = \phi\Lb z; b\Rb$) we obtain the following equation (using the variable $
 \tilde{z}\,\,=\,\sqrt{\frac{16 N_c}{b}}\,z$)
\beq \label{TW3}
 \sqrt{\frac{16 N_c}{b}}\frac{ \tilde{z}}{ \sqrt{2\,Y}}\,\frac{d^2 \phi\Lb \tilde{z}; b \Rb }{d\, \tilde{z}^2}\,\,+\,\,\frac{d^2 \phi\Lb \tilde{z}; b \Rb}{d\, \tilde{z}^2}\,\,=\,\,1 \,
\,-\,\,e^{ - \phi\Lb \tilde{z}; b \Rb}
\eeq
Therefore, only in the vicinity of the critical line where $  \sqrt{\frac{16 N_c}{b}}\, \tilde{z}\,\ll\,\sqrt{2\,Y}$ we can expect the geometric   scaling behaviour of the scattering amplitude. It should be stressed that at large value of $Y$ the region where we have the geometric scaling behaviour becomes rather large. Neglecting the first term in \eq{TW3} we obtain the equation in the same form as for frozen $\as$.  It is easy to find the solution to this equation that satisfies the initial condition of \eq{IBC}.  Actually,  the condition  $  \sqrt{\frac{16 N_c}{b}}\, \tilde{z}\,\ll\,\sqrt{2\,Y}$ can be rewritten as $\as\Lb Q^2_s\Rb\,\ln\Lb r^2Q^2_s\Rb \,\ll\,1$ and it  shows the region in which we can consider the running QCD coupling as being frozen at $r^2=1/Q^2_s$.

\subsection{Self-similar solution}
Generally speaking (see Ref.\cite{MATH} formulae {\bf 3.4.1.1} and {\bf 3.5.2})
\eq{SK3} has a self similar (functional separable) solution $\phi\Lb Y, l;b\Rb \,=\,\phi\Lb \zeta ; b \Rb$ with ( see also \eq{ZETA})
\beq \label{SS1}
\zeta\,\,=\,\,Y\,(l - l_s)
\eeq
For function $\phi\Lb \zeta ; b \Rb$ we can reduce \eq{SK3} to the ordinary differential equation
\beq \label{SS2}
\Lb \zeta\,-\,2\Rb\,\frac{d^2 \phi\Lb \zeta ; b \Rb}{d \zeta^2}\,\,+\,\,\frac{\phi\Lb \zeta ; b \Rb}{d \zeta}\,\,=\,\,1\,\,-\,\,e^{- \phi\Lb \zeta ; b \Rb}
\eeq 

\begin{figure}[h]
\centerline{\epsfig{file=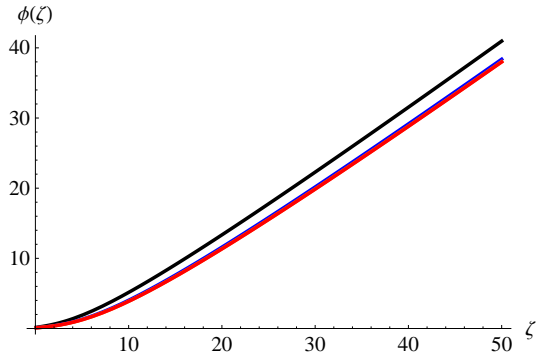,width=70mm}}
\caption{Solution to \protect\eq{SS2} for different values of $ \phi'_\zeta\Lb \zeta=0 ; b \Rb$ (from top to bottom $\phi'_\zeta\Lb \zeta=0 ; b \Rb = 2 $ (black),$ 0.2$ ( blue), $ 0 $ (red)).
\label{zetasc}}
\end{figure}
The initial condition of \eq{IBC} can be rewritten in the form
\bea \label{SSIC}
\phi\Lb \zeta = 0 ; b \Rb\,\,&=&\,\,\phi_0;\,\,\,\,\,\,\,\,\,\,\,\\
\frac{d \phi\Lb \zeta=0 ; b \Rb}{ d \zeta}\,\,&=&\,\,-\frac{3}{4} \phi_0/\xi_s \,\,=\,\,-
\frac{3}{4} \phi_0\Big{/} \sqrt{\frac{32 N_c}{b} \,Y}\nn
\eea

{}

Generally speaking we cannot satisfy  \eq{SSIC} using the solution of \eq{SS2} since these conditions depend  not only on $\zeta$ but on extra variable $Y$.
However, at large value of $Y$ one can see that  \eq{SSIC} degenerates to
\beq \label{SSIC1}
\phi\Lb \zeta = 0 ; b \Rb\,\,=\,\,\phi_0;\,\,\,\,\,\,\,\,\,\,\,\frac{d \phi\Lb \zeta=0 ; b \Rb}{ d \zeta}\,\,=\,\,0;
\eeq

which are consistent with the solution being the function of only $\zeta$. In \fig{zetasc} we plot the numerical solution to the \eq{SS2} at different values of $  \phi'_\zeta\Lb \zeta=0 ; b \Rb$. One can see that this solution is not sensitive to this value if it is small enough. It means that the $\zeta$ scaling behaviour can start from rather small values of $Y$. Since the whole approach, based on leading log(1/x) contribution, can be trusted only at large values of $Y$ we believe that $\zeta$ scaling behaviour is a good approximation to the solution of \eq{SK3}.

\subsection{Numerical solution}
It turns out that this believe was too optimistic. In \fig{exact} we plot the numerical solution to \eq{SK3} with the initial condition given by \eq{IBC}. The main lesson that we can learn from these pictures is that the $\zeta$- scaling behaviour can be reasonable approach but at unreasonably large values of $l$ or/and $Y$. Indeed, at \fig{exact}-b we can see that at large values of $\zeta$ the solition 
depends on $l$ only slowly.

\FIGURE[h]{
\begin{tabular}{c c}
\epsfig{file=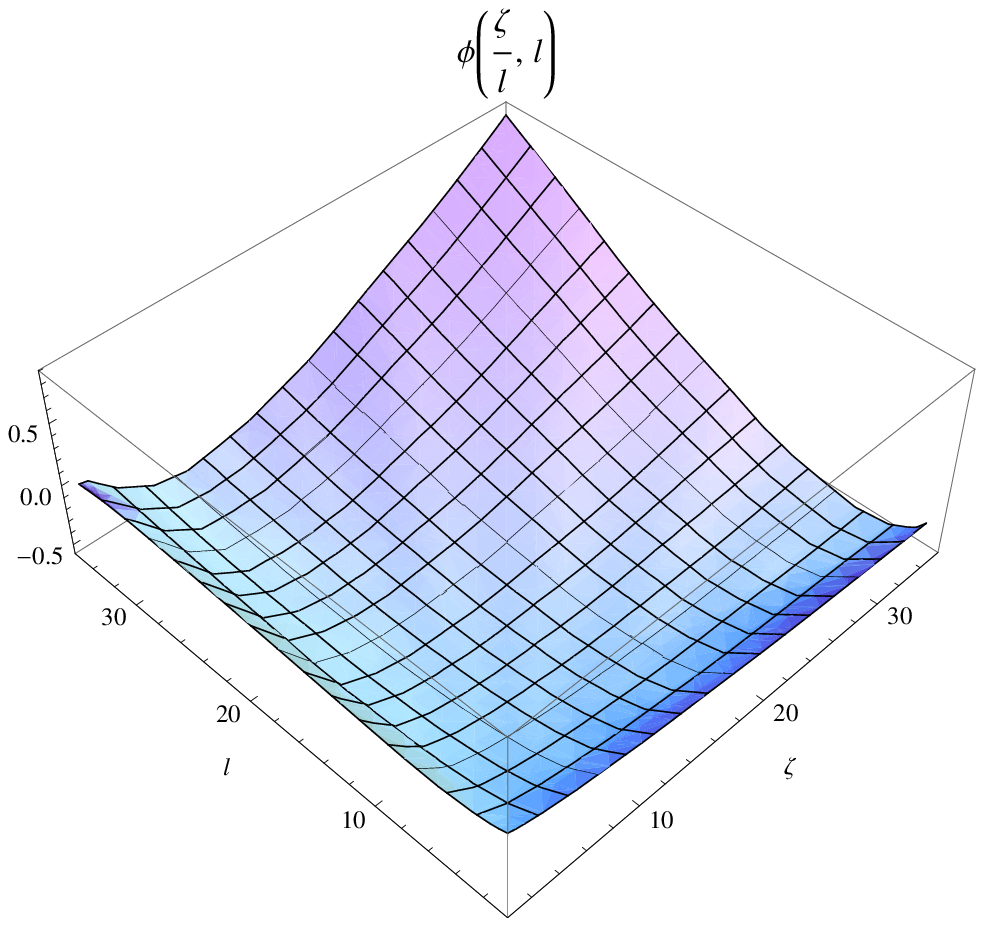,width=80mm} &
\epsfig{file=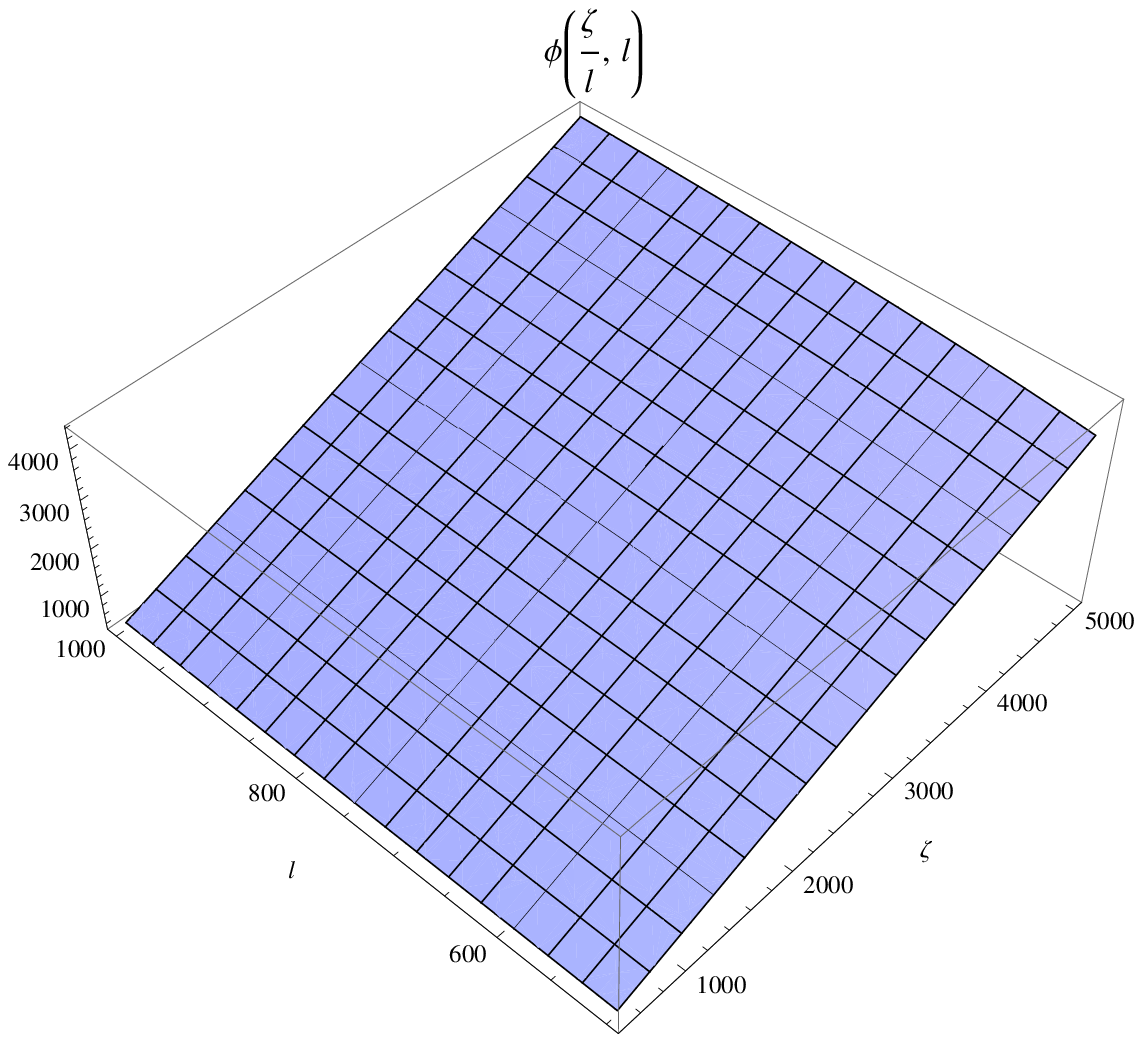,width=75mm}\\
\fig{exact}-a & \fig{exact}-b \\
\end{tabular}
\caption{The exact solution of \eq{SK3}  with the initial conditions given by \eq{IBC}  for function $\phi\Lb \zeta/l,l \Rb$. \fig{exact}-a gives the bahaviour of $\phi$ at small values of $\zeta$ and $l$ while \fig{exact}-b shows the same behaviour in the region of large $\zeta$ and $l$. The value of $\phi_0 $ was taken $\phi_0 = 0.1$
 }
\label{exact}
}
At  large $Y$ and $l$ the exact solution is reasonable to compare with the solution of \eq{SS2} (see \fig{exactvinfty}). One can see the same pattern: they become close at large values of $Y$ and $l$ ($\zeta$ and $l$).


\FIGURE[ht]{
\begin{tabular}{c c}
\epsfig{file=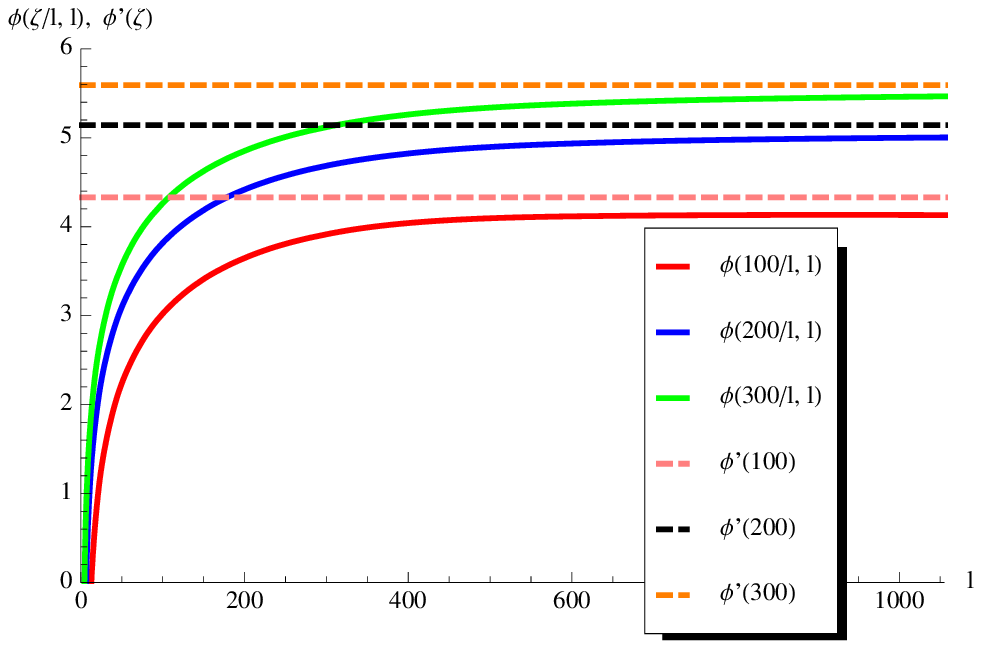,width=75mm} &
\epsfig{file=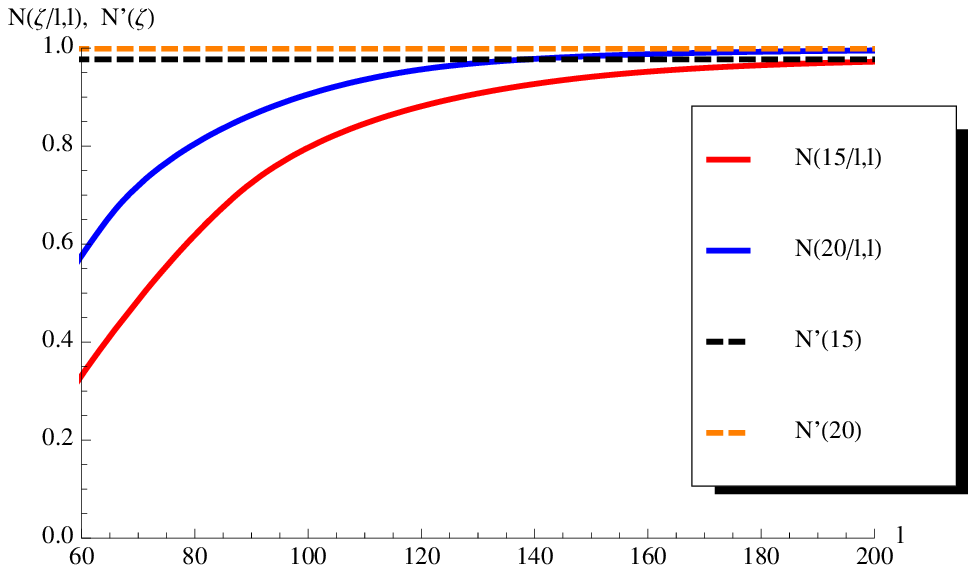,width=80mm}\\
\fig{exactvinfty}-a & \fig{exactvinfty}-b \\
\end{tabular}
\caption{The exact solution of \protect\eq{SK3} with the initial conditions given by \eq{IBC}  for function $\phi\Lb Y, l \Rb$ versus 
$\zeta$-scaling solution of \eq{SS2}  $ \phi'\Lb \zeta \Rb$. In \fig{exactvinfty}-a $\ln \phi\Lb \zeta/l, l \Rb$ and $\ln \phi'\Lb\zeta \Rb$ are plotted at diffrent values of $\zeta$.  \fig{exactvinfty}-b shows the same but for the amplitude $N\Lb \zeta/l,l \Rb = 1 - \exp\Lb -\phi\Lb \zeta/l, l \Rb \Rb$ and  
 $N'\Lb Y\zeta\Rb  = 1 - \exp\Lb - \phi'\Lb \zeta \Rb\Rb$. The calculations were performed at $\phi_0 = 0.1$.
}
\label{exactvinfty}
}

\section{Conclusions}

In this paper we show that the intuitive guess that the running QCD coupling will violate the geometric behaviour of the scattering amplitude, turns out to be correct.  Indeed, we found out that  the new dimensional scale: $\Lambda_{QCD}$, that brings the running $\as$,  enters to the amplitude behaviour  even at very high energies. However, in the vicinity of the saturation scale ($ r^2\, \varpropto \,1/Q^2_s$) we see the geometric scaling behaviour. This vicinity is determined by $\as(Q_s)\,\ln\Lb r^2 Q^2_s\Rb \,\leq \,1$. In other words, the geometric scaling behaviour of the amplitude remains until we can neglect the difference between $\as\Lb r^2\Rb$ and $ \as\Lb 1/Q^2_s\Rb$.  In different way of saying, if we could find the mechanism that will freeze the running $\as$ on $r^2 = 1/Q^2_s$ the amplitude would show the geometric scaling behavior. However, in the framework of the leading twist BFKL we did not find such a mechanism.

For $\as(Q_s)\,\ln\Lb r^2 Q^2_s\Rb \,> \,1$ the geometric scaling behaviour is violated and at very large $Y$ and $l = \ln \Big( \ln\Lb r^2 \Lambda^2_{QCD}\Rb/\ln\Lb Q^2_s \Lambda^2_{QCD}\Rb\Big)$ the amplitude depends only on one variable $\zeta$ (see \eq{ZETA}).

From our point of view the fact that we see geometric scaling behavior experimetally, stems from either we have not probed the proton at HERA and the LHC  deeply inside the saturation region or that there exists the mechanism of  freezing of the coupling QCD constant at $r^2 \approx 1/Q^2_s$.  In practical calculations  $\as$ is used to be frozen at some value of the momentum larger that $1/r_{fr}^2 \,>\,\Lambda^2_{QCD}$. In this case we would like to notice 
that if $\as(Q_s)\,\ln\Lb r^2_{fr} Q^2_s\Rb \,\leq \,1$ we still have the geometric scaling behaviour. For example in recent paper of Ref.\cite{AAMAS}
the value of $r_{fr}$ is chosen from $\as(r_{fr}) = 0.7$ or $1$.  For such value of $r_{fr}$ we  see that   $\as(Q_s)\,\ln\Lb r^2_{fr} Q^2_s\Rb \,\leq \,1$ for
$Q^2_s \,=\,0.3 \div 4\,GeV^2$ covering the region of energy from RHIC to LHC. Therefore, in the CGC motivated model of Ref.\cite{AAMAS} we do not expect to see any violation of the geometric scaling behaviour.

\section*{Acknowledgements}


\end{document}